\documentclass[11pt,a4paper]{article}

\usepackage{epsfig,amsmath,amssymb,cite}

\begin{document}

\title{Dilaton thin--shell wormholes supported by\\
a generalized Chaplygin gas} 
\author{Cecilia Bejarano$^{1,2}$\thanks{e-mail: cbejarano@iafe.uba.ar}, Ernesto F. Eiroa$^{1,3}$\thanks{e-mail: eiroa@iafe.uba.ar}\\
{\small $^1$ Instituto de Astronom\'{\i}a y F\'{\i}sica del Espacio, C.C. 67, 
Suc. 28, 1428, Buenos Aires, Argentina}\\
{\small $^2$ Departamento de Ciencias Exactas, Ciclo B\'asico Com\'un,} \\ 
{\small Universidad de Buenos Aires, Ciudad Universitaria Pab. III, 1428, 
Buenos Aires, Argentina}\\
{\small $^3$ Departamento de F\'{\i}sica, Facultad de Ciencias Exactas y 
Naturales,} \\ 
{\small Universidad de Buenos Aires, Ciudad Universitaria Pab. I, 1428, 
Buenos Aires, Argentina}}
\maketitle

\begin{abstract}
In this article, we construct spherical thin--shell wormholes with charge in dilaton gravity. The exotic matter required for the construction is provided by a generalized Chaplygin gas. We study the stability under perturbations preserving the symmetry. We find that the increase of the coupling between the dilaton and the electromagnetic fields reduces the range of the parameters for which stable configurations are possible.\\

\noindent 
PACS number(s): 04.20.Gz, 04.40.Nr, 98.80.Jk

\end{abstract}

\section{Introduction}\label{intro} 
Since the seminal paper by Morris and Thorne \cite{mothor} traversable Lorentzian wormholes have attracted great interest. As solutions of the equations of gravity, these objects describe geometries which have two regions, either in the same universe or in different universes, connected by a throat \cite{mothor, visser}. In static spacetimes, it is possible to define the throat as a minimal area surface satisfying a flare-out condition \cite{hochvis97}. Within the framework of General Relativity, traversable wormholes must be threaded by exotic matter that violates the null energy condition \cite{mothor, visser, hochvis97, hochvis98}. The amount of exotic matter required at the throat can be made infinitesimally small by appropriately choosing the geometry of the wormhole \cite{viskardad}. Although it may imply that the throat should suffer large stresses \cite{eisi05,zas}. Discussions about the energy conditions in the context of wormholes can be find in Refs.~\cite{barvis02-ro}.

Thin--shell wormholes are mathematical constructions made by cutting and pasting two manifolds to form a geodesically complete one, with a shell located at the joining surface, which corresponds to the throat where the exotic matter is present \cite{visser, vis89}. The junction-condition formalism is used for their study \cite{:junction:}. The stability analysis, under perturbations that preserve the symmetries, has been widely performed for a variety of thin--shell wormholes. Poisson and Visser carried out a linear analysis of the stability of wormholes constructed by joining two equal Schwarzschild geometries \cite{poisvis}. Barcel\'o and Visser studied wormholes built using branes with negative tensions \cite{barvis00}. Ishak and Lake analyzed transparent spherically symmetric thin--shells and wormholes \cite{ishla}. Eiroa and Romero extended the linearized stability study to Reissner-Nordstr\"{o}m wormhole spacetimes \cite{eiro}. Lobo and Crawford did the same analysis with wormhole geometries with a cosmological constant \cite{locraw04}. Thin--shell wormholes were studied within the context of dilaton gravity in Refs.~\cite{eisi05,ei08}. Cylindrical thin--shell wormholes have been also analyzed in Refs.~\cite{:cyl-sym:}. For other recent interesting articles see Refs.~\cite{:recent:}.

The requirement of matter that violates the energy conditions is not only a feature of wormholes. If General Relativity is assumed as the gravity theory to describe the large scale of the Universe, the accelerated expansion violates the strong energy condition (see Refs.~\cite{:obs-data:} for observational data analysis). Therefore several models of exotic matter, which have been proposed in the cosmological context \cite{:cosmo:}, have been also used for wormholes. Phantom energy has been considered in a variety of wormhole articles. A Chaplygin gas \cite{:chaply-orig:,:chaply-cosmo:} was used as the exotic matter in wormhole spacetimes \cite{lo06,eisi07,:chaply-wh:,ei09,:chaply-wh-n:}. In particular, a generalized Chaplygin gas within a finite region around the throat, has been adopted by Lobo \cite{lo06} as the exotic matter supporting wormholes, and by Eiroa \cite{ei09} for thin--shell wormholes, in which the exotic matter is located at the joining surface. 

In the present paper, we construct and analyze the stability under perturbations preserving the symmetry of thin--shell wormholes supported by a generalized Chaplygin gas in dilaton gravity. That is, wormholes constructed by cutting and pasting geometries associated to a charged black holes with dilaton and Maxwell fields. In Sec.~\ref{sec:tswh}, we review the basic aspects of construction and stability of thin--shell wormholes with a generalized Chaplygin gas. In Sec.~\ref{sec:dil-wh}, we perform the study of thin--shell wormholes in dilaton gravity. In Sec.~\ref{sec:conclu}, the results are summarized. Units such that $c=G=1$ are adopted throughout this work.

\section{Basic equations}\label{sec:tswh}
In this Section, we review the formalism developed in Ref.~\cite{ei09} for a wide class of spherically symmetric thin--shell wormholes with a generalized Chaplygin gas. We start from the general metric
\begin{equation} 
ds^2=-f(r)dt^2+f(r)^{-1}dr^2+h(r) (d\theta ^2+\sin^2\theta d\varphi^2), 
\label{e1}
\end{equation}
where $r>0$ is the radial coordinate, $0\le \theta \le \pi$ and $0\le \varphi<2\pi $ are the angular coordinates, and the functions $f(r)$ and $h(r)$ are positive from a given radius. From this geometry, we take two identical copies of a region with radius $r \geq a$ assuming $a >\mathrm{max}\{r_{h},r_{s}\}$ in order to avoid the presence of horizons (at $r_{h}$) and/or singularities (at $r_{s}$): $\mathcal{M}^{\pm }=\{X^{\alpha }=X^{\alpha }(t,r,\theta,\varphi)/r\geq a\}$, and we paste them at the hypersurface $\Sigma \equiv \Sigma ^{\pm }=\{X/F(r)=r-a=0\}$, to create a new manifold $\mathcal{M}=\mathcal{M}^{+}\cup \mathcal{M}^{-}$. If $h'(a)>0$ (condition of flare-out), this construction creates a geodesically complete manifold representing a wormhole with two regions connected by a throat of radius $a$, where the surface of minimal area is located. In such a way that the area of a surface at fixed radius is $\mathcal{A}=4 \pi h(r)$, with $h(r)$ an increasing function for $r \in [a,a+\delta)$, being $\delta > 0$. Thus the area has a minimum at $r=a$ and the manifold $\mathcal{M}$ represents a wormhole with throat $\Sigma$. On this manifold it is possible to define a new radial coordinate $l=\pm \int_{a}^{r}\sqrt{1/f(r)}dr$ (the signs correspond, respectively, to $\mathcal{M}^{+}$ and $\mathcal{M}^{-}$) that represents the proper radial distance to the throat, which is situated at $l=0$. On the wormhole throat, which is a synchronous timelike hypersurface, we  define coordinates $\xi ^{i}=(\tau ,\theta,\varphi )$, with $\tau $ the proper time on the shell. 

For the stability analysis, under perturbations preserving the symmetry, we let the radius of the throat be a function of proper time $\tau$. Outside the throat, the geometry remains static. The lack of gravitational waves is guarantee by the Birkhoff theorem (see Refs.~\cite{bronkov-bronmel} for the conditions that should be satisfied in a spherically symmetric geometry). Following the Darmois-Israel formalism \cite{:junction:}, the second fundamental forms (or extrinsic curvature) associated with the two sides of the shell are given by
\begin{equation} 
K_{ij}^{\pm }=-n_{\gamma }^{\pm }\left. \left( \frac{\partial ^{2}X^{\gamma
} } {\partial \xi ^{i}\partial \xi ^{j}}+\Gamma _{\alpha \beta }^{\gamma }
\frac{ \partial X^{\alpha }}{\partial \xi ^{i}}\frac{\partial X^{\beta }}{
\partial \xi ^{j}}\right) \right| _{\Sigma },  
\label{e4}
\end{equation}
where $n_{\gamma }^{\pm }$ are the unit normals ($n^{\gamma }n_{\gamma }=1$) to $\Sigma $ in $\mathcal{M}$:
\begin{equation} 
n_{\gamma }^{\pm }=\pm \left| g^{\alpha \beta }\frac{\partial F}{\partial
X^{\alpha }}\frac{\partial F}{\partial X^{\beta }}\right| ^{-1/2}
\frac{\partial F}{\partial X^{\gamma }}.  
\label{e5}
\end{equation}
Using the orthonormal basis $\{ e_{\hat{\tau}}=e_{\tau }, e_{\hat{\theta}}=[h(a)]^{-1/2}e_{\theta }, e_{\hat{\varphi}}=[h(a)\sin^2 \theta ]^{-1/2} e_{\varphi }\} $, we obtain for the geometry (\ref{e1}) that
\begin{equation} 
K_{\hat{\theta}\hat{\theta}}^{\pm }=K_{\hat{\varphi}\hat{\varphi}}^{\pm
}=\pm \frac{h'(a)}{2h(a)}\sqrt{f(a)+\dot{a}^2},
\label{e6}
\end{equation}
and
\begin{equation} 
K_{\hat{\tau}\hat{\tau}}^{\pm }=\mp \frac{f'(a)+2\ddot{a}}{2\sqrt{f(a)+\dot{a}^2}},
\label{e7}
\end{equation}
where a prime and the dot represent, respectively, the derivatives with respect to $r$ and $\tau$. Defining $[K_{_{\hat{\imath}\hat{\jmath}}}]\equiv K_{_{\hat{\imath}\hat{\jmath}}}^{+}-K_{_{\hat{\imath}\hat{\jmath}}}^{-}$, $K=tr[K_{\hat{\imath}\hat{\jmath }}]=[K_{\; \hat{\imath}}^{\hat{\imath}}]$ and introducing  the surface stress-energy tensor $S_{_{\hat{\imath}\hat{\jmath} }}={\rm diag}(\sigma ,p_{\hat{\theta}},p_{\hat{\varphi}})$, where $\sigma$ is the surface energy density and $p_{\hat{\theta}} = p_{\hat{\varphi}} = p$ are the transverse pressures, the Einstein equations on the shell (the well-known Lanczos equations) are
\begin{equation} 
-[K_{\hat{\imath}\hat{\jmath}}]+Kg_{\hat{\imath}\hat{\jmath}}=8\pi 
S_{\hat{\imath}\hat{\jmath}},
\label{e8}
\end{equation}
so we get 
\begin{equation} 
\sigma=-\frac{\sqrt{f(a)+\dot{a}^2}}{4\pi }\frac{h'(a)}{h(a)},
\label{e9}
\end{equation}
and
\begin{equation}
p=p_{\hat{\theta}}=p_{\hat{\varphi}}=\frac{\sqrt{f(a)+\dot{a}^2}}{8\pi}\left[ \frac{2\ddot{a}+f'(a)}{f(a)+\dot{a}^2}+\frac{h'(a)}{h(a)}\right] .
\label{e10}
\end{equation}
The negative sign in Eq. (\ref{e9}) plus the flare-out condition ($h'(a)>0$) implies that $\sigma <0$, indicating the presence of exotic matter at the throat.

We adopt a generalized Chaplygin gas as the exotic matter in the shell $\Sigma $. For this gas, the pressure has opposite sign to the energy density, resulting in a positive pressure. Then, the equation of state for the exotic matter at the throat can be written in the form
\begin{equation}
p=\frac{A}{|\sigma |^{\alpha}},
\label{e11} 
\end{equation} 
where $A>0$ and $0<\alpha\le 1$ are constants. The original Chaplygin gas  \cite{:chaply-orig:} with equation of state $p=-A/\sigma$ is recovered for $\alpha=1$, and it has the property that the squared sound speed is always positive, even in the case of exotic matter. Then, by replacing Eqs. (\ref{e9}) and (\ref{e10}) in Eq. (\ref{e11}), we obtain the differential equation that should be satisfied by the throat radius of thin--shell wormholes threaded by exotic matter with the equation of state of a generalized Chaplygin gas:
\begin{equation}
\{ [2\ddot{a}+f'(a)]h(a)+[f(a)+\dot{a}^2]h'(a) \}[h'(a)]^{\alpha}-2A[4\pi h(a)]^{\alpha +1}[f(a)+\dot{a}^2]^{(1-\alpha)/2}=0.
\label{e12} 
\end{equation}
For the static solutions with throat radius $a_{0}$, the surface energy density and pressure are given by
\begin{equation}
\sigma_{0}=-\frac{\sqrt{f(a_{0})}}{4\pi}\frac{h'(a_{0})}{h(a_{0})},
\label{e13}
\end{equation}
and
\begin{equation}
p_{0}=\frac{\sqrt{f(a_{0})}}{8\pi}\left[ \frac{f'(a_{0})}{f(a_{0})}+\frac{h'(a_{0})}{h(a_{0})}\right] .
\label{e14}
\end{equation}
The throat radius of the static solutions, $a_{0}$, should fulfill Eq. (\ref{e12}), that is
\begin{equation}
[f'(a_{0})h(a_{0})+f(a_{0})h'(a_{0})][h'(a_{0})]^{\alpha}-2A[4\pi h(a_{0})]^{\alpha +1}[f(a_{0})]^{(1-\alpha)/2}=0,
\label{e15} 
\end{equation} 
with the condition $a_{0}>$max$\{r_{h},r_{s}\}$ (if the metric has an event horizon or a singularity). 

Following the standard potential approach, we study the stability of the static solution, under perturbations that preserve the symmetry. From Eqs. (\ref{e9}) and (\ref{e10}) (and recalling that the area of the wormhole throat is $\mathcal{A}=4\pi h(a)$), the energy conservation equation is found:
\begin{equation}
\frac{d}{d\tau }\left( \sigma \mathcal{A}\right) +p\frac{d\mathcal{A}}{d\tau }=
\left\lbrace \left[ h'(a)\right]^2 -2h(a)h''(a)\right\rbrace \frac{\dot{a}\sqrt{f(a)+ \dot{a}^2}}{2h(a)}.
\label{p1}
\end{equation}
The first term in the left hand side is the internal energy change of the throat and the second the work done by the internal forces of the throat, while the right hand side represents a flux. If $[h'(a)]^2 -2h(a)h''(a)=0$, the flux term is zero and Eq. (\ref{p1}) takes the form of a simple conservation equation. This occurs when $h(a)=C(a+D)^2$  (with $C>0$ and $D$ constants) or  $h(a)=C$  (this case is unphysical, since there is no throat \cite{ei09}). It is straightforward to see that Eq. (\ref{p1}) can be written in the form
\begin{equation}
h(a)\sigma '+h'(a)(\sigma +p)+\left\lbrace \left[ h'(a)\right]^2 -2h(a)h''(a)\right\rbrace
\frac{\sigma }{2 h'(a)}=0,
\label{p3}
\end{equation}
where it was used that $\sigma '=\dot{\sigma }/\dot{a}$. Since the pressure $p$ is a function of $\sigma $ given by the equation of state, Eq. (\ref{p3}) is a first order differential that can be recast in the form $\sigma '(a)=\mathcal{F}(a, \sigma (a))$, for which a unique solution with a given initial condition always exists, provided that $\mathcal{F}$ has continuous partial derivatives. Then, Eq. (\ref{p3}) can be formally integrated to obtain $\sigma (a)$.  If we replace $\sigma (a)$ in Eq. (\ref{e9}), the dynamics of the wormhole throat is completely determined by a single equation:
\begin{equation}
\dot{a}^{2}=-V(a),
\label{p4}
\end{equation}
with 
\begin{equation}
V(a)=f(a)-16\pi ^{2}\left[ \frac{h(a)}{h'(a)}\sigma (a)\right] ^{2}.
\label{p5}
\end{equation}
Replacing the surface energy density and the pressure corresponding to the static solutions (Eqs. (\ref{e13}) and (\ref{e14})) and using that $V(a_{0})=V'(a_{0})=0$, a Taylor expansion to second order of the potential $V(a)$ around the static solution yields \cite{ei09}:
\begin{equation}
V(a)=\frac{1}{2}V''(a_{0})(a-a_{0})^{2}+O[(a-a_{0})^{3}],
\label{p11}
\end{equation}
with 
\begin{eqnarray}
V''(a_{0})&=& f''(a_{0})+\frac{(\alpha -1)[f'(a_{0})]^2}{2f(a_{0})}+\left[ \frac{(1-\alpha )h'(a_{0})}{2h(a_{0})}+\frac{\alpha h''(a_{0})}{h'(a_{0})}\right] f'(a_{0}) \nonumber \\
& & +(\alpha +1)\left[ \frac{h''(a_{0})}{h(a_{0})} -\left( \frac{h'(a_{0})}{h(a_{0})} \right) ^{2}\right] f(a_{0}).
\label{p12}
\end{eqnarray}
As usual, if the dynamics of the system is completely determined by an equation like Eq. (\ref{p4}), the stability under radial perturbations is guaranteed if and only if $V''(a_{0})>0$. 
\section{Dilaton wormholes}\label{sec:dil-wh}
The four-dimensional modified Einstein action in dilaton gravity includes the (scalar) dilaton field $\phi$ and the electromagnetic field $F^{\mu\nu}$, and  in the Einstein frame has the form \cite{:dil-grav:}
\begin{equation} S=\int d^4x\sqrt{-g}\left(-R+2(\nabla \phi)^2+e^{-2b\phi}F^2\right),
\label{emd}
\end{equation}
where $R$ is the Ricci scalar of the spacetime. The parameter $b$ represents the coupling between the dilaton and the Maxwell field and it will be restricted within the range $0 \leq b \leq 1$. For $b=0$, the action corresponds to the usual Einstein--Maxwell scalar theory. For $b=1$, the action was obtained in the context of low energy string theory with a Maxwell field, but with all other gauge fields and the antisymmetric field set to zero. The action given by Eq.~(\ref{emd}) leads to the Einstein equations with the dilaton and the Maxwell fields as the sources \cite{:dil-grav:}:
\begin{equation}
\nabla _{\mu }\left( e^{-2b\phi }F^{\mu \nu }\right) =0,
\end{equation}
\begin{equation}
\nabla ^{2}\phi +\frac{b}{2}e^{-2b\phi }F^{2}=0,
\end{equation}
\begin{equation}
R_{\mu \nu }=2\nabla _{\mu }\phi \nabla _{\nu }\phi +2e^{-2b\phi }
\left( F_{\mu \alpha }F_{\nu }^{\; \alpha }-\frac{1}{4}g_{\mu \nu }F^{2}\right) .
\label{feq}
\end{equation}
Assuming $\phi = \phi(r)$, which means that the dilaton field does not depend on time, these field equations admit spherically symmetric black hole solutions \cite{:dil-grav:,gibmae} in the  form of Eq. (\ref{e1}), with metric functions, in Schwarzschild coordinates, given by 
\begin{eqnarray}
f(r) & = &\left( 1-\frac{A}{r}\right)\left( 1-\frac{B}{r}
\right)^{(1-b^2)/(1+b^2)}, \label{d1a}\\
h(r) & = &r^2\left( 1-\frac{B}{r}\right)^{2b^2/(1+b^2)},
\label{d1b}
\end{eqnarray}
where the constants $A,B$ are related with the parameter $b$, and the mass and charge of the black hole by
\begin{equation}
A=M+\sqrt{M^2-(1-b^2)Q^2},
\label{d3a}
\end{equation} 
and
\begin{equation}
B=\frac{(1+b^2)Q^2}{M+\sqrt{M^2-(1-b^2)Q^2}}.
\label{d3b}
\end{equation}
In the case of electric charge, the electromagnetic field tensor has non-null components $F_{tr}=-F_{rt}=Q/r^{2}$, and the dilaton field is given by $\phi =b(1+b^2)^{-1}\ln( 1-B/r)$, where the asymptotic value of the dilaton $\phi_{0}$ was taken as zero. For magnetic charge, the metric is the same, with the electromagnetic field $F_{\theta \varphi}=-F_{\varphi \theta}=Q\sin \theta $ and the dilaton field obtained replacing $\phi $ by $-\phi $. 

We can distinguish three cases: i) $b=0$, ii) $0< b < 1$, and iii) $b=1$. In the first case, the geometry corresponds to Reissner-Nordstr\"om metric, with $f(r)=1-2M/r+Q^2/r^2$ and $h(r)=r^{2}$. If $|Q|/M< 1$, the constants $A$ and $B$ represent, respectively, the radii of the outer (event) and the inner horizons; if $|Q|/M=1$ both horizons have the same radius; and if $|Q|/M>1$ there is only a naked singularity at $r_s=0$.  In the second case, the metric has an event horizon for $r_{h}=A$ and a singular surface for $r_s=B$, and there are two possibilities: if $0 \leq |Q|/M \leq (1+b^{2})^{1/2}$, then $A \geq B$ which implies that $r_{h} \geq r_{s}$; if $ (1+b^{2})^{1/2} < |Q|/M \leq (1-b^{2})^{-1/2}$, we have $A < B$ and the event horizon is located inside the singularity radius (so called naked singularity). It is worthy of mention that there is a maximum value for the charge: $(1-b^{2})^{-1/2}$, otherwise the geometry is not well defined. In the third case, one obtains $f(r)=1-2M/r$ and $h(r)=r^{2}[1-Q^2/(2Mr)]$. The event horizon at $r_{h}=A=2M$ is outside  or inside the singularity at $r_{s}=B=Q^2/(2M)$, accordingly to $|Q| \leq \sqrt{2}M$  or $|Q| > \sqrt{2}M$, respectively.
                                                                                                                                                                                                                                   
Reissner-Nordstr\"om wormholes ($b=0$) with a generalized Chaplygin gas were previously studied by one of the authors \cite{ei09}. In the present work, we focus on dilaton wormholes ($ 0< b \le 1$). As it was done in a previous work \cite{eisi05}, we construct thin--shell wormholes in dilaton gravity, assuming that the throat has a radius $a$ greater than $A$ and $B$ to eliminate the presence of horizons and singularities. For the geometry represented by the metric (\ref{e1}), the static solutions with throat radius $a_{0}$ are obtained replacing $f(a_{0})$ and $h(a_{0})$ in Eqs. (\ref{e13}) and (\ref{e14}), thus we reobtain the surface energy density and pressure found in Ref. \cite{eisi05}
\begin{equation}
\sigma _{0}=-\frac{1}{2\pi a_{0}^2}\left( 1-\frac{A}{a_{0}}\right)^{1/2}\left( 1-\frac{B}{a_{0}}
\right)^{(1-b^2)/(2+2b^2)}\left[ a_{0}+\frac{b^2B}{1+b^2}\left( 1-\frac{B}{a_{0}}
\right)^{-1}\right],
\label{d4}
\end{equation}
\begin{equation}
p_{0}=\frac{1}{8\pi a_{0}^{2}}\left( 1-\frac{A}{a_{0}}\right) ^{1/2}\left( 1-\frac{B}{a_{0}}
\right) ^{(1-b^2)/(2+2b^2)}\left[ 2a_{0}+A\left( 1-\frac{A}{a_{0}}\right) ^{-1}+B
\left( 1-\frac{B}{a_{0}}\right) ^{-1}\right].
\label{d5}
\end{equation}
But now the pressure and the surface energy density are related by the generalized Chaplygin gas equation of state, instead of the linearized equation of state adopted in Ref.~\cite{eisi05}. It is straightforward to see that the surface energy density and pressure corresponding to the Reissner--Nordstr\"om thin--shell wormholes studied in Ref.~\cite{ei09} are recovered when $b=0$.
\begin{figure}[t!]
\begin{center}
\vspace{-1cm}
\includegraphics[width=15cm]{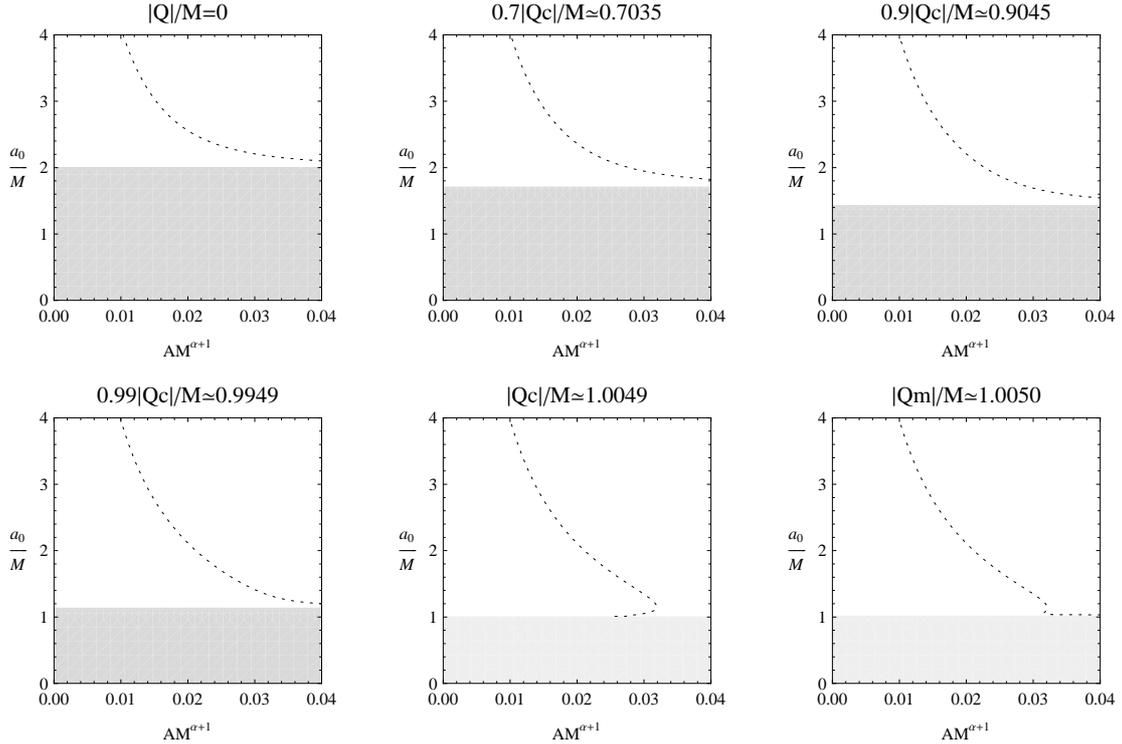}
\vspace{-0.5cm}
\end{center} 
\caption{Dilaton wormholes with $b=0.1$ supported by a generalized Chaplygin gas with $\alpha =0.2$: the dashed curves represent the unstable static solutions with throat radius $a_{0}$ for given parameters $A$, $M$ and $Q$; there are not stable static solutions. The darker (lighter) gray zones are unphysical, corresponding to a throat radius smaller than the horizon (singularity) radius of the original manifold. The event horizon and the naked singularity match for a critical value of charge $|Q_{c}|/M=(1+b^{2})^{1/2}$ and the geometry is not well defined for $|Q|/M>|Q_{m}|/M=(1-b^{2})^{-1/2}$.}
\label{f1}
\end{figure} 
\begin{figure}[t!]
\begin{center}
\vspace{-1cm}
\includegraphics[width=15cm]{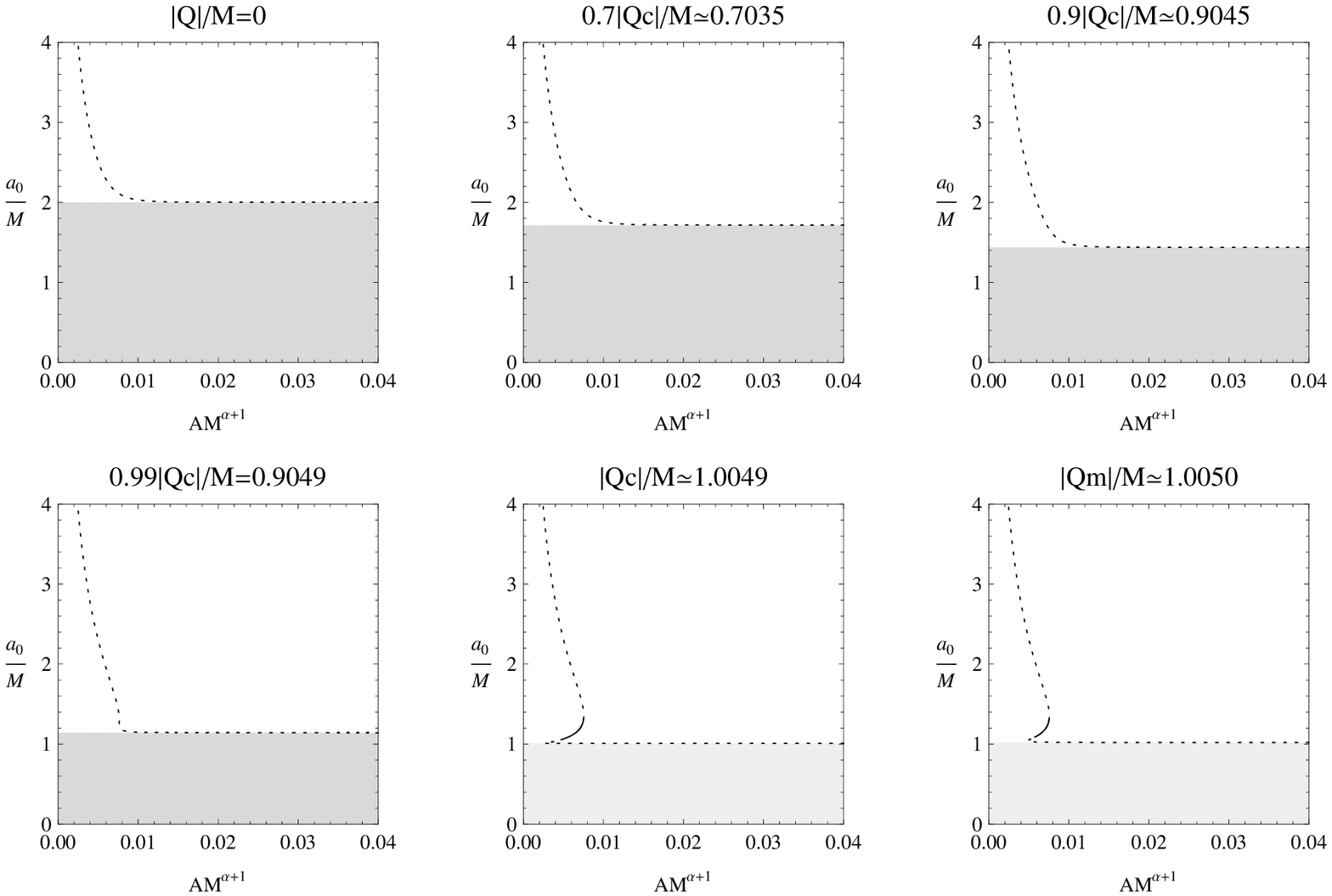}
\vspace{-0.5cm}
\end{center} 
\caption{Dilaton wormholes with $b=0.1$ supported by a generalized Chaplygin gas with $\alpha =0.6$: the dashed curves represent the unstable static solutions with throat radius $a_{0}$ for given parameters $A$, $M$ and $Q$; the solid curves represent the stable static solutions. The meaning of the gray zones and the charges $Q_c$ and $Q_m$ are explained in Fig. \ref{f1}.}
\label{f2}
\end{figure}
\begin{figure}[t!]
\begin{center}
\vspace{-1cm}
\includegraphics[width=15cm]{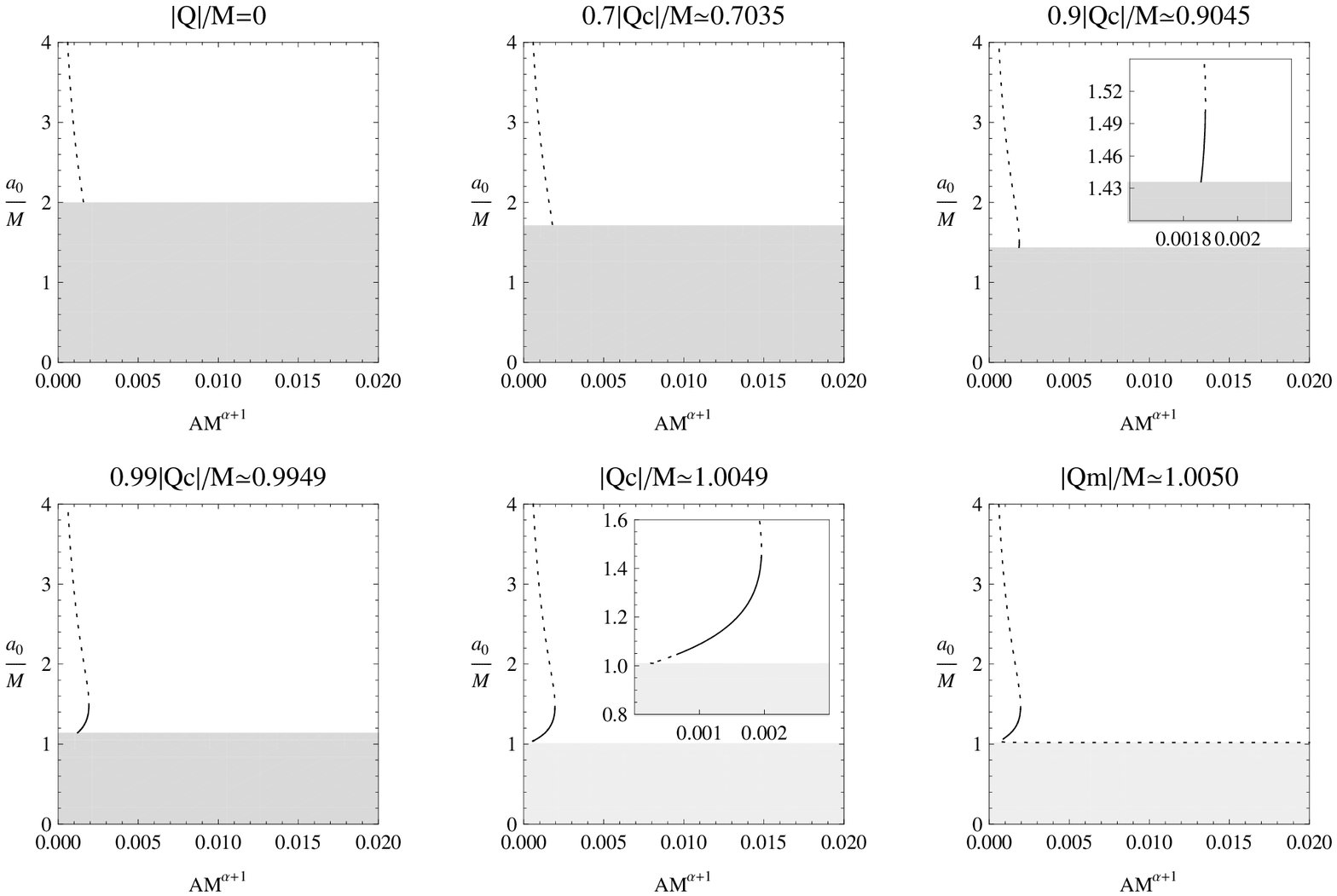}
\vspace{-0.5cm}
\end{center} 
\caption{Dilaton wormholes with $b=0.1$ supported by a generalized Chaplygin gas with $\alpha =1$: the dashed curves represent the unstable static solutions with throat radius $a_{0}$ for given parameters $A$, $M$ and $Q$; the solid curve represent the  stable static solutions. The meaning of the gray zones and the charges $Q_c$ and $Q_m$ are explained in Fig. \ref{f1}. Note that the horizontal scale has changed.}
\label{f3}
\end{figure}
\begin{figure}[t!]
\begin{center}
\vspace{-1cm}
\includegraphics[width=15cm]{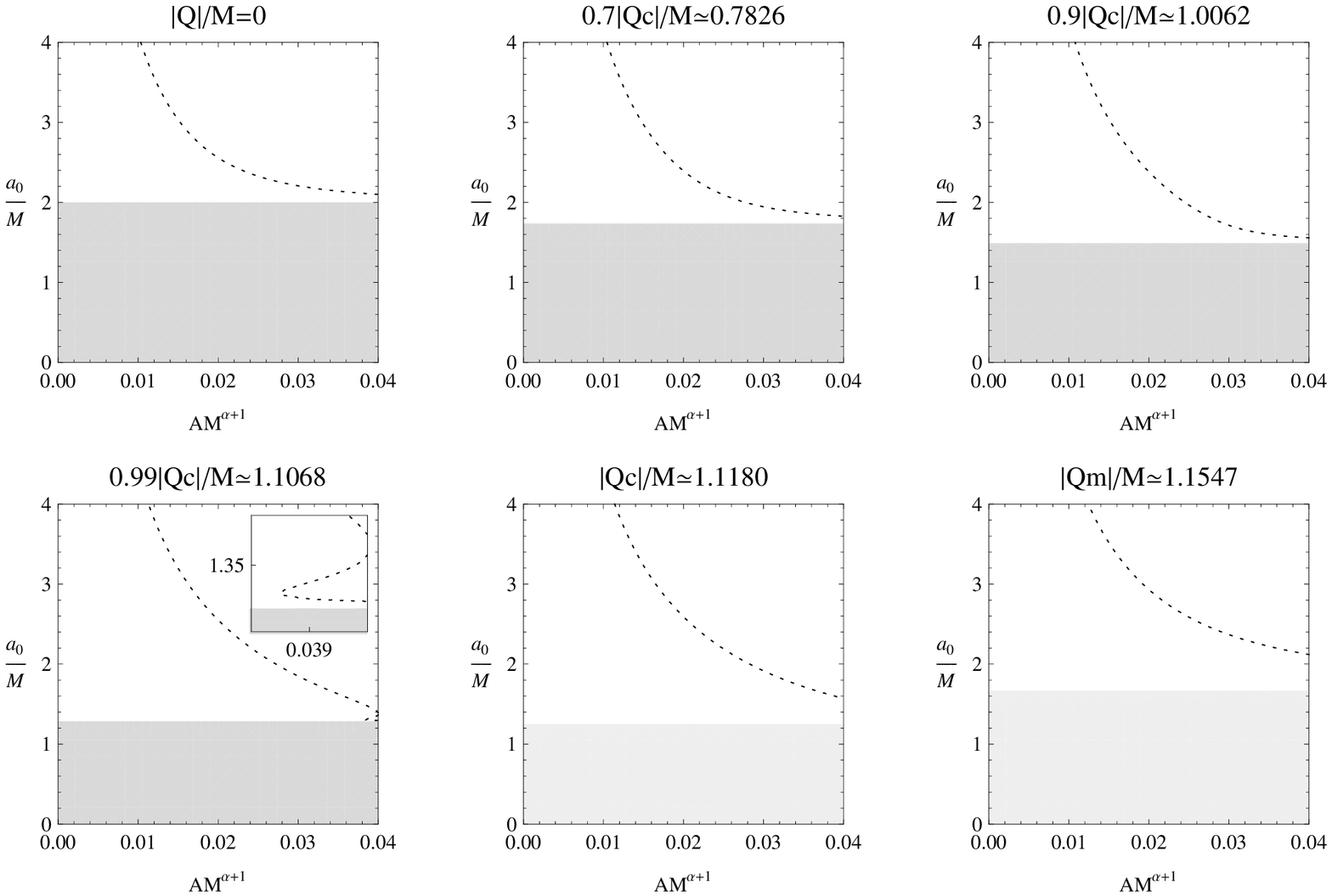}
\vspace{-0.5cm}
\end{center} 
\caption{Dilaton wormholes with $b=0.5$ supported by a generalized Chaplygin gas with $\alpha =0.2$: the dashed curves represent the unstable static solutions with throat radius $a_{0}$ for given parameters $A$, $M$ and $Q$; there are not stable static solutions. The meaning of the gray zones and the charges $Q_c$ and $Q_m$ are explained in Fig. \ref{f1}.}
\label{f4}
\end{figure} 
\begin{figure}[t!]
\begin{center}
\vspace{-1cm}
\includegraphics[width=15cm]{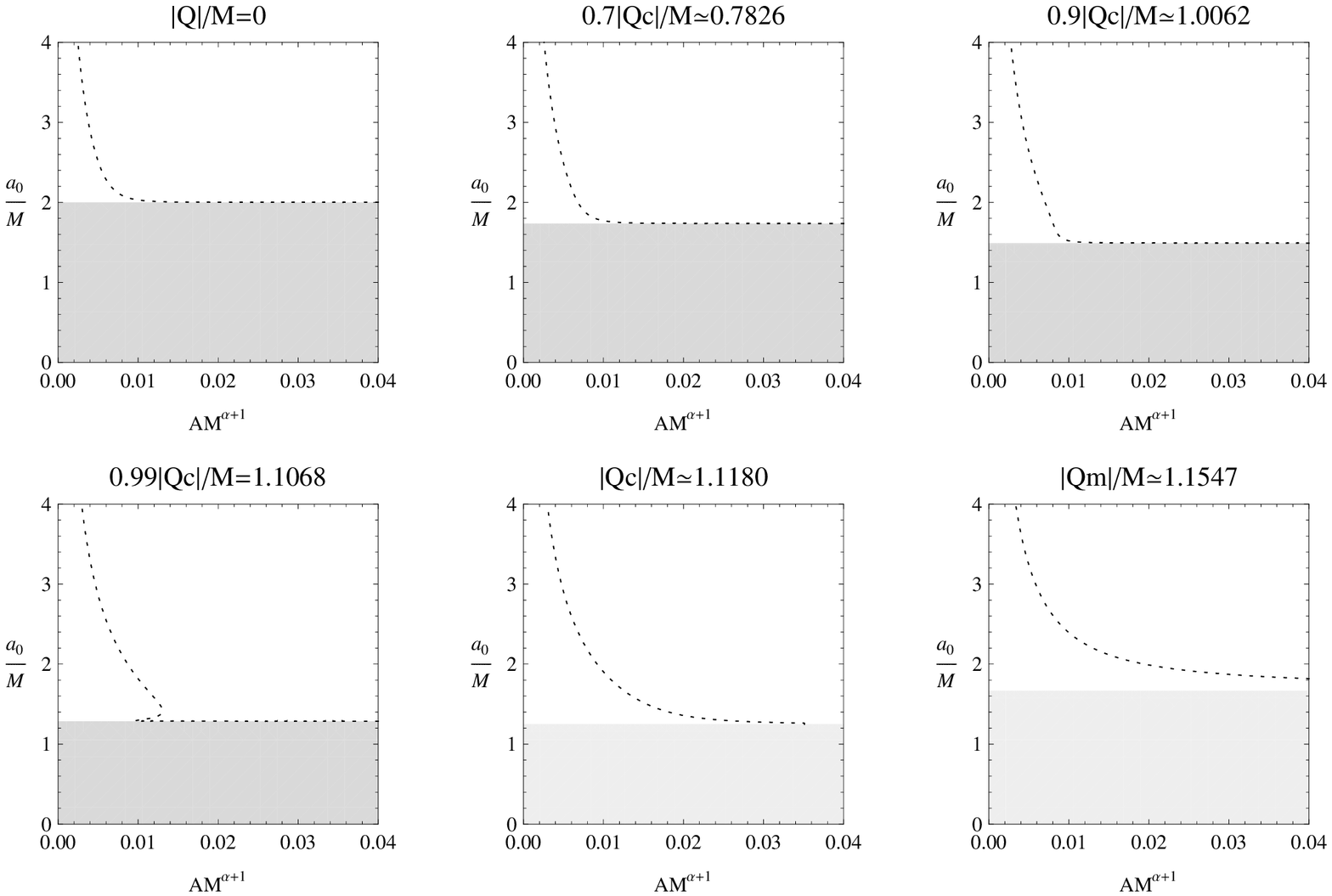}
\vspace{-0.5cm}
\end{center} 
\caption{Dilaton wormholes with $b=0.5$ supported by a generalized Chaplygin gas with $\alpha =0.6$: the dashed curves represent the unstable static solutions with throat radius $a_{0}$ for given parameters $A$, $M$ and $Q$; there are not stable static solutions. The meaning of the gray zones and the charges $Q_c$ and $Q_m$ are explained in Fig. \ref{f1}.}
\label{f5}
\end{figure}
\begin{figure}[t!]
\begin{center}
\vspace{-1cm}
\includegraphics[width=15cm]{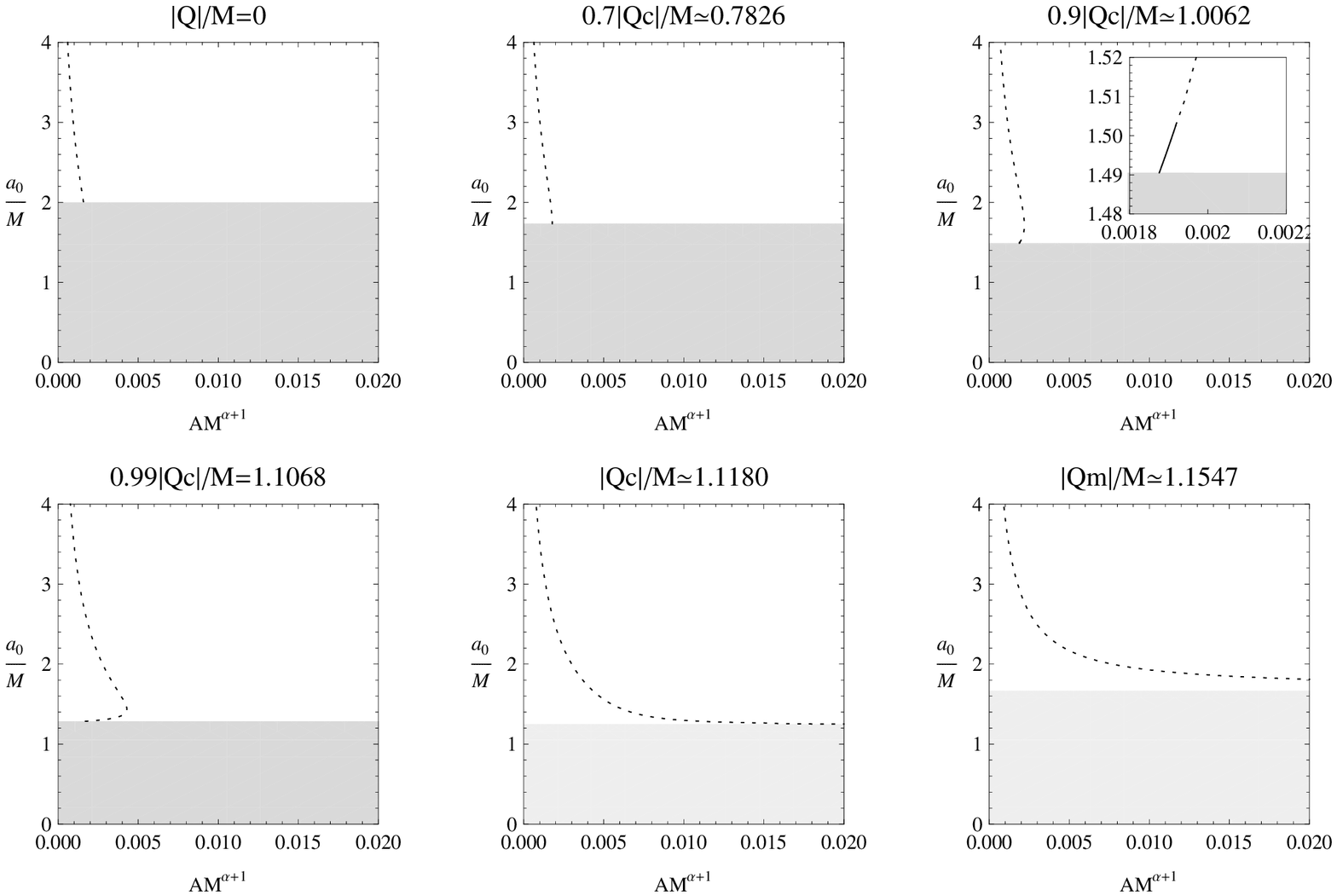}
\vspace{-0.5cm}
\end{center} 
\caption{Dilaton wormholes with $b=0.5$  supported by a generalized Chaplygin gas with $\alpha =1$: the dashed curves represent the unstable static solutions with throat radius $a_{0}$ for given parameters $A$, $M$ and $Q$; the solid curve represent the  stable static solutions (see the magnified region). The meaning of the gray zones and the charges $Q_c$ and $Q_m$ are explained in Fig. \ref{f1}. Note that the horizontal scale has changed. }
\label{f6}
\end{figure}

When $0 < b < 1$, we assume that the throat radius satisfy that  $a_{0}>A>B$ for $0 \leq |Q| < |Q_{c}|$, since the event horizon contain the singularity, and $a_{0}>B \geq A$ for $|Q_{c}| \leq |Q| < |Q_{m}|$, because of the event horizon is inside of the singular surface (thus is a naked singularity); we have labeled the critical value of charge, for which the event horizon and the singularity have the same radial coordinate, as $|Q_{c}|/M=(1+b^{2})^{1/2}$. As we pointed out above, the charge cannot take values greater than what we have called $|Q_{m}|/M=(1-b^{2})^{-1/2}$ as the maximum possible value of charge for which the geometry is well defined.  In the particular case of $b=1$ there is no maximum value of charge since the geometry is always well defined in our wormhole construction. 

The throat radii of the static solutions are obtained by replacing Eqs. (\ref{d1a}) and (\ref{d1b}) in Eq.~(\ref{e15}), and solving it numerically; these solutions are stable when satisfy that the second derivative of the potential, given by Eq.~(\ref{p12}), is positive. For a given set of parameters, the complexity of the expression of $V''(a_{0})$ compel us to find the range of $a_0$ for which $V''(a_{0})>0$  in a numerical form and display the results graphically. In Figs.~\ref{f1}--\ref{f3} we show the results obtained for values of dilaton parameter $b=0.1$ which is close to the Reissner-Nordstr\"om case (i.e.\ $b=0$), and $\alpha =0.2$, $\alpha =0.6$, $\alpha =1$, respectively. In Figs.~\ref{f4}--\ref{f6} we display the results found for $b=0.5$, corresponding to an intermediate value of dilaton parameter, and $\alpha =0.2$, $\alpha =0.6$, $\alpha =1$, respectively. The values of charge adopted were chosen to be representative in each case. In all plots, the solid curves correspond to stable solutions while the dashed lines to unstable ones.

As shown in Fig.~\ref{f1} (where $b=0.1$ and $\alpha =0.2$), for small values of $\alpha$ there are not stable solutions. This is a very important difference with the Reissner-Nordstr\"om wormholes, studied in Ref.~\cite{ei09}. We can summarize the results as follows:
\begin{itemize}
\item When $0 \leq |Q| < |Q_{c}|$ there is one unstable solution with a throat radius $a_{0}/M$ which decreases with $AM^{\alpha +1}$ and approaches to the event horizon radius of the original manifold for large values of $AM^{\alpha+1}$.
\item If $|Q| = |Q_{c}|$ there is one unstable solution for small values of $AM^{\alpha +1}$, there are two unstable solutions for intermediate values of $AM^{\alpha +1}$, and there are no solutions for large values of $AM^{\alpha +1}$, because the throat radius $a_{0}/M$ reaches the singularity of the original manifold at a finite value of $AM^{\alpha +1}$.
\item When $|Q| > |Q_{c}|$ there is one unstable solution for small values of $AM^{\alpha +1}$, there are three unstable solutions for intermediate values of $AM^{\alpha +1}$, and there is one unstable solution for large values of $AM^{\alpha +1}$, with the throat radius $a_{0}/M$ approaching asymptotically to the singularity of the original manifold. 
\end{itemize}  

As shown in Fig.~\ref{f2} (where $b=0.1$ and $\alpha =0.6$), for intermediate values of $\alpha$ there are stable solutions for properly chosen values of the parameters. The main results are:
\begin{itemize}
\item For $0 \leq |Q| < |Q_{c}|$  there is one unstable solution with a throat radius $a_{0}/M$ that asymptotically approaches to the radius of the event horizon of the original manifold for large values of $AM^{\alpha +1}$, faster than in the case of the small values of $\alpha$. 
\item For $|Q| \geq |Q_{c}|$ there are three solutions at small values of $AM^{\alpha +1}$: the larger and the smaller ones  are unstable and the middle one is stable, while for large values of $AM^{\alpha +1}$ there is only one unstable solutions with a throat radius $a_{0}/M$ that tends asymptotically to the singularity of the original manifold.
\end{itemize}
Note that all cases have solution for any value of $AM^{\alpha +1}$.

As shown in Fig.~\ref{f3} (where $b=0.1$ and $\alpha =1$), the case of the ordinary Chaplygin gas presents stable solutions. We find that:
\begin{itemize}
\item When $0 \le |Q| < |Q_{c}|$ and $|Q|$ is not very close to $|Q_{c}|$, for small values of $AM^{\alpha +1}$ there is only an unstable solution. The throat radius $a_{0}/M$ decreases with $AM^{\alpha +1}$ and it cuts the horizon radius of the original manifold at a finite value of $AM^{\alpha +1}$, so there are no solutions for intermediate and large values of $AM^{\alpha +1}$.
\item If $|Q|<|Q_{c}|$ and $|Q|$ is very close to $|Q_{c}|$, for small values of $AM^{\alpha +1}$ there are two solutions with throat radius  $a_{0}/M$: the larger one is unstable and the smaller one is stable.  The throat radius cuts the horizon radius of the original manifold at a finite and very close to zero value of $AM^{\alpha +1}$,  so there are no solutions for intermediate and large values of $AM^{\alpha +1}$.
\item If $|Q|=|Q_{c}|$, for small values of $AM^{\alpha +1}$ there are three solutions with throat radius  $a_{0}/M$: the larger and the smaller ones are unstable, and the intermediate one is stable. For intermediate and large values of $AM^{\alpha +1}$ there are no solutions since the throat radius reaches the singularity of the original manifold at very small values of $AM^{\alpha + 1}$.
\item When $|Q_{c}|<|Q|<|Q_{m}|$, for small values of $AM^{\alpha +1}$ there are three solutions with throat radius  $a_{0}/M$: the larger and the smaller are unstable, the middle one is stable; while for intermediate and large values of $AM^{\alpha +1}$ there is only one unstable solution that asymptotically approaches to the singularity of the original manifold.
\end{itemize}

In Fig.~\ref{f4} (where $b=0.5$ and $\alpha =0.2$), the solutions are unstable for all values of charge. The throat radius is a decreasing function of $AM^{\alpha +1}$ that presents asymptotic behavior for large values of $AM^{\alpha +1}$. Sufficiently closer to the critical value of charge, there are three unstable solutions for a very small interval of  $AM^{\alpha +1}$.  In Fig.~\ref{f5} (where $b=0.5$ and $\alpha =0.6$), for all values of charge the solutions are unstable. The throat radius approaches asymptotically to the event horizon  of the original manifold or to the singularity of the original manifold (depending on the value of charge), in a faster way than the previous case, with a small value of $\alpha$. Sufficiently closer to the critical value of charge, there are three unstable solutions for a narrow interval of  $AM^{\alpha +1}$. For both, $\alpha =0.2$ and $\alpha =0.6$, there are solutions for any value of $AM^{\alpha +1}$. 

In Fig.~\ref{f6} (where $b=0.5$ and $\alpha =1$) there is one stable configuration, and the main results are the following:
\begin{itemize}
\item When $0 \le |Q| < |Q_{c}|$ and $|Q|$ is not very close to $|Q_{c}|$, for small values of $AM^{\alpha +1}$ there is only an unstable solution. The throat radius $a_{0}/M$ decreases with $AM^{\alpha +1}$ and it cuts the horizon radius of the original manifold at a finite value of $AM^{\alpha +1}$, so there are no solutions for intermediate and large values of $AM^{\alpha +1}$.
\item If $|Q|<|Q_{c}|$ and $|Q|$ is very close to $|Q_{c}|$, but not too much close, for small values of $AM^{\alpha +1}$ there are two solutions with throat radius $a_{0}/M$: the larger one is unstable and the smaller one is stable. The throat radius reaches the horizon radius of the original manifold for a finite and very close to zero value of $AM^{\alpha +1}$,  so there are no solutions for intermediate and large values of $AM^{\alpha +1}$. Note that the horizontal scale was reduced in the plots in order to obtain a better visualization of the little region of the stability.
\item When $|Q|<|Q_{c}|$ and $|Q|$ is near enough to $|Q_{c}|$ there are two unstable solutions for small values of $AM^{\alpha +1}$. The throat radius cuts the horizon radius of the original manifold at a finite and very close to zero value of $AM^{\alpha +1}$,  so there are no solutions for intermediate and large values of $AM^{\alpha +1}$.
\item If $|Q| \geq |Q_{c}|$, there is only one unstable solution with a throat radius that tends asymptotically to the singularity of the original manifold.
\end{itemize}
When $b=0.5$ and $0 < \alpha < 1$, we have found that there are not stable solutions. Only the case $\alpha = 1$, that is the ordinary Chaplygin gas, presents stable configurations.

In all cases analyzed, when there are both stable and unstable static configurations, the throat radius $a_{0}/M$ of the stable solution is an increasing function of $AM^{\alpha +1}$, while the unstable branch decreases with $AM^{\alpha +1}$. As the value of the dilaton parameter increases, we find that the interval of $a_{0}/M$ for which $V''(a_{0})>0$ reduces. We observe that there are no roots of $V''(a_{0})$ for $0< \alpha <1$ when the dilaton parameter is such that $b\gtrsim 0.22$.  So that the second derivative remains negative for all possible values of $AM^{\alpha+1}$, which means that the dilaton wormholes have no stable static solutions for $0< \alpha <1$ if $b\gtrsim  0.22$. In the specific case of $\alpha=1$, which corresponds to the ordinary Chaplygin gas, we obtain a very small stable region for some values of the charge close enough to $Q_{c}$  (but not too much close). The interval of $a_{0}/M$ for which there are stable static solutions is smaller as far as the value of the dilaton coupling parameter increases. In particular, for the uppermost value $b=1$, we encountered a very small interval, for a very narrow range of values of charge, where the wormholes have stable behavior (the plots are not shown here for brevity).

\newpage

\section{Conclusions}\label{sec:conclu}
In this work, we have studied the stability under perturbations preserving the spherical symmetry of thin--shell wormholes in dilaton gravity supported by a generalized Chaplygin gas. For the construction we have performed the usual cut and paste procedure; so the surface energy density and the pressure at the throat have been obtained following the junction formalism. For the stability analysis we have implemented the standard potential approach. We have found that for a fixed value of the dilaton coupling parameter $b$, the possibility of obtaining a static stable solution increases as far as the exponent $\alpha$ in the equation of state grows, i.e. as the generalized Chaplygin gas gets closer to the ordinary one (with $\alpha = 1$). The same result was previously found for Reissner--Nordstr\"om  thin--shell wormholes (which correspond to $b=0$) in Ref.~\cite{ei09}. On the other hand, as the dilaton coupling $b$ increases, the values of the parameters for which stable static solutions exist lie in a smaller range, so the dilaton field reduces the presence of stable configurations.

\section*{Acknowledgments}

This work has been supported by CONICET and Universidad de Buenos Aires.

\end{document}